\RequirePackage{fix-cm}
\documentclass[twocolumn]{svjour3}          
\smartqed  
\usepackage{graphicx}
\usepackage{amsmath}
\usepackage{graphicx,psfrag,epsf}
\usepackage{enumerate}
\usepackage{natbib}
\usepackage{graphicx}			
\usepackage{amssymb,amsmath,bm}
\usepackage{mathtools}
\usepackage{commath}
\usepackage[english]{babel}
\usepackage{hyperref}
\usepackage{threeparttable}
\usepackage{enumerate}
\usepackage{float}
\usepackage{bbm}
\usepackage{graphicx} 
\usepackage{booktabs} 
\usepackage{bm}
\setcitestyle{authoryear,open={(},close={)}}

\usepackage{algorithm,algcompatible,amsmath}
\algnewcommand\INPUT{\item[\textbf{Input:}]}%
\algnewcommand\OUTPUT{\item[\textbf{Output:}]}%

\usepackage{tikz,xcolor,hyperref}
\definecolor{lime}{HTML}{A6CE39}
\DeclareRobustCommand{\orcidicon}{%
	\begin{tikzpicture}
	\draw[lime, fill=lime] (0,0) 
	circle [radius=0.16] 
	node[white] {{\fontfamily{qag}\selectfont \tiny ID}};
	\draw[white, fill=white] (-0.0625,0.095) 
	circle [radius=0.007];
	\end{tikzpicture}
	\hspace{-2mm}
}

\foreach \x in {A, ..., Z}{%
	\expandafter\xdef\csname orcid\x\endcsname{\noexpand\href{https://orcid.org/\csname orcidauthor\x\endcsname}{\noexpand\orcidicon}}
}

\DeclareMathOperator{\Var}{Var}

\begin{document}

\title{Kernel-based Approximate Bayesian Inference for Exponential Family Random Graph Models
\thanks{This work was supported in part by National Science Foundation (NSF) awards IIS-1526736, DMS-1361425, and IIS-1939237.}
}

\titlerunning{K-ABC for ERGMs}        
\author{Fan Yin \orcidA{} \and
        Carter T. Butts 
}


\institute{           Fan Yin \at
              Department of Statistics, University of California, Irvine, CA 92697, USA \\
              \email{yinf2@uci.edu}  
              \and
Carter T. Butts \at
              Departments of Sociology, Statistics, and EECS, and Institute for Mathematical Behavioral Sciences, University of California, Irvine, CA 92697, USA \\
              \email{buttsc@uci.edu}           
}

\date{Received: date / Accepted: date}

\maketitle

\begin{abstract}
Bayesian inference for exponential family random graph models (ERGMs) is a doubly-intractable problem because of the intractability of both the likelihood and posterior normalizing factor. Auxiliary variable based Markov Chain Monte Carlo (MCMC) methods for this problem are asymptotically exact but computationally demanding, and are difficult to extend to modified ERGM families. In this work, we propose a kernel-based approximate Bayesian computation algorithm for fitting ERGMs. By employing an adaptive importance sampling technique, we greatly improve the efficiency of the sampling step. Though approximate, our easily parallelizable approach is yields comparable accuracy to state-of-the art methods with substantial improvements in compute time on multi-core hardware. Our approach also flexibly accommodates both algorithmic enhancements (including improved learning algorithms for estimating conditional expectations) and extensions to non-standard cases such as inference from non-sufficient statistics. We demonstrate the performance of this approach on two well-known network data sets, comparing its accuracy and efficiency with results obtained using the approximate exchange algorithm. Our tests show a wallclock time advantage of up to 50\% with five cores, and the ability to fit models in 1/5th the time at 30 cores; further speed enhancements are possible when more cores are available. 
\end{abstract}

\section{Introduction}
\label{sec:intro}

Exponential family random graph models (ERGMs) \citep{holland1981exponential, frank1986markov, snijders2006new, hunter2006inference}, also known as $p$-star models\citep{wasserman1996logit}, are a family of parametric statistical models representing the complex stochastic processes that govern the formation of edges between pairs of nodes in a network. ERGMs have found wide application in many scientific fields, for example, including (but not limited to) sociology \citep{srivastava2011culture, smith2016ethnic}, political science \citep{cranmer2011inferential}, biology \citep{saul2007exploring,grazioli2019abeta}, public health \citep{o2013analysis} and neuroscience \citep{simpson2011exponential, sinke2016bayesian}. 

Despite remarkable success in network modeling, parametric inference for ERGMs with complex dependence has been a historical challenge and continues to offer open problems for current research.  The central challenge stems from the normalizing factor of the ERGM likelihood, which involves integrating an extremely rough function over all possible network configurations. While somewhat ad-hoc methods based on path lengths were explored in a pre-ERGM context by e.g. \citet{rapoport1957contribution,fararo1964study,rapoport1979probabilistic}, the first work to investigate inference for random graphs with dependence structure in a fully modern sense was the iterative scaling algorithm proposed for the $p_1$ model \citep{holland1981exponential}, now identified as a sub-class of ERGMs where the dependence is within each dyad (i.e. reciprocity). As an attempt to incorporate higher-order dependence structure, \citet{frank1986markov} introduced the \emph{Markov graphs}, where edge variables are dependent only if they share a common node; unfortunately, the accompanying estimation algorithm based on cumulant approximations was not practical for use in typical settings. A major advance was made with Strauss and Ikeda's (\citeyear{strauss1990pseudolikelihood}) adaptation of the maximum pseudolikelihood estimation (MPLE) strategy of \citet{besag1974spatial}), in which the likelihood is approximated by a product of full conditional distributions, to the estimation of ERGMs. MPLE is still in use to date in some applications, being relatively fast, algorithmically convenient, and able to provide parameter estimates (albeit sometimes innaccurate ones) for even badly-specified models.  As an approximation to the MLE, however, the MPLE is often biased with respect to the mean value parameter space (which the MLE is not), less efficient than the MLE, prone to instability, and very poorly calibrated \citep{van2009framework}.   Given these issues, most subsequent work has focused on attempting to perform maximum likelihood estimation (MLE).  Fitting general ERGMs using maximum likelihood is numerically demanding, as the likelihood can only be specified up to a parameter dependent normalizing constant, making the exact calculation of the MLE extremely difficult except for extremely small graphs \citep{von.et.al:sn:2020} or in cases for which the likelihood function can be analytically simplified (e.g. homogeneous and inhomogeneous  Bernoulli graphs). State-of-the-art frequentist estimation approaches for ERGMs thus hinge on simulation-based algorithms to obtain high-quality approximations to the MLE, including  \emph{Markov chain Monte Carlo maximum likelihood estimation} (MCMC MLE), originally introduced by \citep{geyer1992constrained} and adapted to ERGMs by \citet{handcock2003assessing, hunter2006inference} and \emph{stochastic approximation} (SA), originally introduced by \citet{robbins1951stochastic, pflug1996optimization} and adapted to ERGMs by \citet{snijders2002markov}. Bayesian inference for general ERGMs is even more challenging and has been historically prohibitive, as the parameter dependent normalizing constant in the likelihood does not cancel out when taking posterior ratios (as is required e.g. for standard MCMC strategies). This produces a target distribution which is termed \emph{doubly intractable} \citep{Murray:2006:MDD:3020419.3020463}, given the intractability of both the likelihood and the normalizing constant of the posterior density, and rendering conventional sampling schemes (e.g. Metropolis-Hastings algorithms \citep{metropolis1953equation, hastings1970monte}) impractical. There have been some recent developments on asymptotic approximations for ERGMs \citep{pu2015deterministic,he2015glmle}, but 
they are derived for a very specific set of models employing only permutation invariant subgraph statistics, typically do not converge in the sparse graph regime, and cannot be employed for models with covariate effects or other inhomogeneities.

The development of Bayesian inference has the potential to offer special advantages vis a vis several issues arising in typical ERGM use cases. Per standard theories of exponential-family models \citep{Barndorff-Nielsen1978}, the MLE for an ERGM's parameters does not exist (i.e., no finite maximizer of the likelihood exists) when the observed statistics for a given model happen to lie on the relative boundary of the convex hull of possible values of the sufficient statistics \citep{handcock2003assessing}.  This issue is not peculiar to ERGMs, and indeed is present in all discrete exponential families (including trivial cases like the binomial model).  However, typical ERGM specifications often include statistics based on sums of small numbers of sparse binary variables, creating a high risk of observing at least one extreme statistic.  Though this can be partially ``patched'' by taking the estimate to be the infinite limit of the parameter in the direction of recession, the resulting model is overconfident (e.g., it may predict that ties between two groups not observed to be in contact are not only rare, but impossible) and lacks well-defined standard errors.  By contrast, Bayes estimators under suitably regular priors will are still well-defined in such cases, and will shrink estimates away from extreme values.  As another example, the standard error of the MLE is currently obtained by employing the inverse of the Fisher information matrix, an approach that is conventionally justified by asymptotics under replication. In typical use cases, however, models are based on only one observed graph, raising questions about the appropriateness of the underlying theory.  While recent results have provided positive justification for using such approaches for certain classes of ERGMs \citep{schweinberger2019foundations}, and there is empirical evidence showing that the resulting estimates of standard error are similar to that yielded by parametric bootstrap \citep{fellows2012exponential}, it is attractive to have alternative frameworks for quantifying uncertainty that do not depend on asymptotic assumptions. Bayesian answers regarding uncertainty in parameter estimates are well-defined even in the finite sample case, and hence provide an immediate way of addressing this issue that does not depend on any particular model specification. 

As noted, fully Bayesian inference for ERGMs is a doubly intractable problem, with both the likelihood and the normalizing constant of the posterior being infeasible to calculate.  Early attempts at resolving this issue were based on conventional Metropolis-Hastings algorithms in which the likelihood ratio at each iteration is approximated by a linked importance sampler auxiliary variable algorithm \citep{koskinen2008linked, koskinen2010analysing}. \citet{caimo2011bayesian} attempted to improve performance by proposing an approximate version of the \emph{exchange algorithm} \citep{Murray:2006:MDD:3020419.3020463} to draw posterior samples of model parameters. This approximate exchange algorithm has become the state-of-the-art approach to Bayesian inference for ERGMs, with the potential to yield high-quality posterior draws, but the algorithm can be very expensive to use due to the need to serially draw high-quality ERGM simulations in an auxiliary chain at each iteration. \citet{bouranis2017efficient} introduce an approximate method to the approximate exchange sampler by calibrating the posterior samples drawn from a ``pseudo-posterior'' -- where the exact ERGM likelihood is replaced by a tractable approximation (e.g. the pseudo-likelihood) -- via an affine transformation that requires the existence of the mode of likelihood (i.e. MLE) and a Monte-Carlo approximation of the gradient and curvature around the mode of likelihood (i.e. MCMC-MLE if MLE cannot be solved precisely).  Pseudo-posteriors are also employed by \citet{grazioli2019abeta}, who instead obtain posterior draws using a ``Bayesian bootstrap'' procedure; though computationally efficient, this approach is limited to cases in which large numbers of graphs are observed from the same generating process.  As these examples suggest, the existing approaches to Bayesian inference for ERGMs are, broadly speaking, either limited to relatively special cases or computationally expensive, non-parallelizable, and difficult to extend to new settings (e.g., ERGMs with endogenous vertex sets \citep{almquist2014bayesian} or inference from non-sufficient statistics) without substantial re-engineering of the underlying algorithms. This hence remains an area of active research, with substantial room for new techniques. 

In this article, we consider another possible direction for fast Bayesian estimation of ERGM parameters by proposing a parallelizable kernelized approximate Bayesian computation (K-ABC) algorithm. We show that the proposed algorithm can yield comparable 
estimates to the gold-standard approximate exchange sampler, with significantly reduced computational time when multiple cores are available. We discuss the choice of distance measure, kernel functions, bandwidth selection, and offer some guidance on selecting optimal settings. We also show the inherent connection between the proposed algorithm and Kernel Bayes' rule (KBR) \citep{fukumizu2011kernel}, offering an interpretation of the resulting estimates from a kernel regression perspective.  The KBR interpretation provides a more direct route to obtaining estimates of posterior moments, and also suggests the opportunity to exploit developments in machine learning (e.g., kernelized WLS) to obtain improved posterior approximations.

The outline of the remainder of this article is as follows. In Section \ref{sec:intro}, we give an introduction to the exponential random graph models along with some simulation and computational methods that serve as building blocks for the proposed method. In Section \ref{sec:ergm}, we propose our parallelizable kernelized approximate Bayesian computation (K-ABC) algorithm for fast Bayesian estimation of ERGMs, and provide implementation details. In Section \ref{sec:applications}, we describe the application of our approach in the context of two benchmark social network datasets of varying sizes, showing the accuracy and computational efficiency of our algorithm compared to approximate exchange algorithm, which is the current ''gold standard" for Bayesian inference of ERGMs. We discuss possible future extensions of the proposed algorithm in Section \ref{sec:externsions} and conclude in Section \ref{sec:conc}.

%
\section{Exponential family random graph models}
\label{sec:ergm}

Consider a random adjacency matrix $Y$ representing a graph on node set $\mathcal{N} = \{1, 2, \cdots, n \}$, in which each element $Y_{ij} = 1$ or $0$ indicates the presence or absence of a tie between node $i$ and $j$. For simplicity in exposition, we here consider graphs and digraphs without multiplex edges or loops, and let $\mathcal{Y}_{n}$ be the set of all such networks on $n$ nodes. We write the probability mass function (pmf) of $Y$ taking a particular configuration $y$, in the form of exponential family as
\begin{equation} 
\label{eq:ERGM}
p(y|\bm{\theta}) = \exp\left\{ \bm{\theta}^{\intercal} g(y) - \psi( \bm{\theta} ) \right\} 
\end{equation}
where $g : \mathcal{Y}_{n} \rightarrow \mathbb{R}^{p}$ is a vector of user-defined sufficient statistics capturing network features of interest, which may implicitly incorporate covariates that are measured on the nodes or dyads; $\bm{\theta} = (\theta_{1}, \cdots, \theta_{p}) \in \Theta \subseteq \mathbb{R}^{p}$ is the vector of model parameters. In general, computation involving this PMF \eqref{eq:ERGM} is challenging due to the intractable nature of the log-partition function (i.e. normalizing factor),
$\psi(\theta) = \log \sum_{y' \in \mathcal{Y}_{n} } \exp\left\{ \bm{\theta}^\intercal g(y') \right\} $, which involves the summation of  $2^{n \choose 2}$ non-negative terms for any value of $\theta$. Note that $2^{n \choose 2}$ is an astronomically large number even for moderate $n$ (e.g., for n = 15 there are 4.1 $\times 10^{31}$ terms) and hence the direct evaluation of such summation is prohibitive for all but trivially small graphs; more importantly, $\bm{\theta}^{\intercal} g(y)$ is in general an extremely rough function over $\mathcal{Y}_n$, making naive Monte Carlo strategies or other simple approximations ineffective. 

In some cases, it is useful to think of ERGMs in terms of distributions on collections of edge variables instead of on graphs.  Specifically, let $\mathcal{D}$ denote the set of all pairs of dyads on $\mathcal{N}$. For directed networks, $\mathcal{D} = \{ (i,j) | i,j \in \mathcal{N}, i \neq j  \}$, while for undirected networks, $\mathcal{D} = \{ (i,j) | i,j \in \mathcal{N}, i < j  \}$. $Y_{ij}$ can then be viewed as an indicator for the event that the $(i,j)$ element of $\mathcal{D}$ contains an edge.  Note that $|\mathcal{D}| \sim \mathcal{O}(n^2)$, and hence ERGMs provide a compact way of representing the joint distribution of $\mathcal{O}(n^2)$ correlated (binary) random variables.

The development of computational approaches for ERGMs is briefly reviewed in section \ref{sec:intro}; in this section, we focus on the technical details of the methods that serve as building blocks of and points of comparison for the proposed ABC-based algorithm. For additional technical details regarding other conventional computational approaches, we refer the interested readers to the review of \citet{hunter2012computational}. 

\subsection{Maximum pseudolikelihood estimation}
\label{subsec:mple}
An approximate approach to maximum likelihood estimation for ERGMs is based on the \emph{pseudolikelihood} function \citep{strauss1990pseudolikelihood}, 
\begin{equation}
\label{eq:ERGM_PL}
p(y|\bm{\theta}) \approx f_{PL}(y|\bm{\theta}) = \prod_{(i,j) \in \mathcal{D}} p(y_{ij} | y_{-ij}, \bm{\theta}).
\end{equation}
\eqref{eq:ERGM_PL} is simply the product of full conditional distributions, as we can see from the conditional specification of the ERGM PMF,
\begin{align}
\text{logit}\left\{  p(y_{ij} = 1 | y_{-ij}, \bm{\theta})  \right\} &= \log \frac{p(y_{ij} = 1 | y_{-ij}, \bm{\theta})}{p(y_{ij} = 0 | y_{-ij}, \bm{\theta})} \nonumber \\
&= \bm{\theta}^{\intercal} \left\{ g(y_{ij}^{+}) - g(y_{ij}^{-})    \right\} \nonumber \\
&= \bm{\theta}^{\intercal} \Delta_{i,j} g(y) \label{eq:ERGM_change_score}
\end{align}
where $\Delta_{i,j} g(y) = g(y_{ij}^{+}) - g(y_{ij}^{-})$ are the so-called \emph{change statistics} associated with the dyad $(i,j)$, representing the change in sufficient statistics when $y_{ij}$ is toggled from 0 ($y_{ij}^{-}$) to 1 ($y_{ij}^{+}$) with the rest of the network remaining unchanged. Following \eqref{eq:ERGM_PL}, the log pseudo-likelihood can be written as 
\begin{align}\label{eq:ERGM_log_PL}
\log f_{PL}(y|\bm{\theta}) & = \sum_{(i,j) \in \mathcal{D}} \big[ y_{ij} \text{logit}\left\{ p(y_{ij} = 1 | y_{-ij}, \bm{\theta}) \right\} + \nonumber \\
& \ \ \ \ \log \left\{ 1  -  p(y_{ij} = 1 | y_{-ij}, \bm{\theta})  \right\} \big] \nonumber \\
& = \sum_{(i,j) \in \mathcal{D}} \big[ y_{ij} \bm{\theta}^{\intercal} \Delta_{ij} g(y) - \nonumber \\
& \ \ \ \ \log\left\{ 1 + \exp( \bm{\theta}^{\intercal} \Delta_{ij} g(y))  \right\} \big] 
\end{align}
Note that \eqref{eq:ERGM_log_PL} is no different from the likelihood of a logistic regression where $y_{ij}$ are the responses and $\Delta_{ij} g(y)$ as the corresponding row in the model matrix, facilitating fast estimation. For exponential family distributions with log-likelihood $\ell(\bm{\theta})$, the estimate of standard error is based on the inverse of Fisher information matrix $\bm{I}^{-1}(\bm{\theta})$, where 
\begin{equation}
\label{eq:fisher_info}
\bm{I}(\bm{\theta}) = \mathbb{E}_{\bm{\theta}} [ \nabla \ell(\bm{\theta}) \nabla \ell(\bm{\theta}) ^{\intercal} ] = Var_{\bm{\theta}}[ g(Y) ].
\end{equation}
Note that under the pseudolikelihood framework, we substitute $\log f_{PL}(y|\bm{\theta})$ for the true log-likelihood $\ell(\bm{\theta}) \equiv \log p(y|\bm{\theta})$, where $y$ is omitted for the convenience of notation. In fact, pseudolikelihood is a special form of composite likelihood \citep{lindsay1988composite}, which is a more general class of inference functions used to approximate complex likelihoods \citep[see][for a review]{varin2011overview}. Despite the empirical observations that MPLE can be unstable and leads to biased standard errors \citep{van2009framework} (especially for models with strong dyadic dependence), it has been the default choice for the initial value in MCMC MLE. There is also promising work on using bootstrapped MPLE to construct confidence intervals \citep{schmid2017exponential} for large and sparse networks, as the MPLE is usually close to the MLE in such cases \citep{desmarais2010consistent}. 


\subsection{Simulation Methods}
\label{subsec:simu}
More advanced estimation techniques, including simulation-based methods for finding maximum likelihood estimates, as well as Bayesian methods, require sampling from the ERGM distribution. To simulate from $p(y|\bm{\theta})$, \citet{snijders2002markov} proposed to use a Metropolis-Hastings sampling procedure: given a proposal $y'$ from density $q(y' | y)$, accept with probability
\begin{align}\label{eq:ERGM_simulation_MH}
\alpha = & \min \bigg(1, \frac{q(y|y') p(y'|\bm{\theta})}{q(y'|y) p(y|\bm{\theta})} \bigg) \nonumber \\
= & \min \bigg(1, \frac{q(y|y')}{q(y'|y)} \exp\left\{\bm{\theta}^{\intercal} (g(y') - g(y) )   \right\} \bigg).
\end{align}
Based on \eqref{eq:ERGM_change_score}, note that if we restrict $q(y'|y) > 0$ only if $\sum_{(i,j) \in \mathcal{D}} \mathbbm{1} (y'_{ij} \neq y_{ij})$ equals one, i.e. only the networks that can be constructed by toggling exactly one dyad from $y$ are allowed to be proposed, then $g(y') - g(y)$ reduces to $\pm \Delta_{ij} g(y)$ with the sign depending on the direction of the toggle.\footnote{Gibbs sampling is also trivial here, as described by \citet{snijders2002markov}, but is generally less efficient.}  To avoid unnecessary computational cost spent on highly improbable graphs, the starting point $y^{(0)}$ of the sampling is usually set as the observed network $y^{obs}$ (if available). Furthermore, as opposed to the basic MCMC algorithm in which each dyad is selected to be toggled uniformly at random, the adoption of asymmetric proposals have been demonstrated to be more favorable for sparse graphs. For example, the ``TNT" (tie-no-tie) sampler implemented as default in the \texttt{ergm} package for R \citep{morris2008specification}, while the ``OTNT"( open triangle-tie-no tie) \citep{wang2014approximate} has been shown to improve performance in clustered networks, and the improved fixed density (IFD) sampler has been shown to be a promising tool for large, sparse networks \citep{byshkin2016auxiliary}. Specifically, the ``TNT" sampler selects an edge or null to toggle with respective probability 1/2 (instead of the graph density, which is close to 0 for a sparse graph) at each iteration, which often leads to better mixing in the typical case of ERGMs that concentrate probability mass on sparse graphs. The above MCMC routines produce a sequence of networks $\left\{y^{(0)}, \cdots, y^{(T)}  \right\}$, of which the initial part is highly dependent on the starting point (network) and hence is usually discarded as burn-in. These are referred to as the \emph{auxiliary iterations} required before a simulated network can be claimed as a random draw from $p(y|\bm{\theta})$. Exact sampling from ERGMs is also possible at a higher computational cost \citep{butts2018perfect}, and non-MCMC approximate samplers have also been proposed \citep{butts2015novel}.

\subsection{Bayesian inference for ERGMs}
We consider the Bayesian treatment of ERGM inference as introduced by \citet{koskinen2004bayesian}. Given observed network $y^{obs}$, and prior distribution $\pi(\bm{\theta})$ placed on $\bm{\theta}$, the full posterior distribution of $\bm{\theta}$ is 
\begin{align}
\pi(\bm{\theta} | y^{obs}) & = \frac{p(y^{obs}|\bm{\theta}) \pi(\bm{\theta})}{\int p(y^{obs}|
\theta) \pi(\bm{\theta}) d\bm{\theta}} \nonumber \\
& \propto p(y^{obs}|\bm{\theta}) \pi(\bm{\theta}) \label{eq:ergm_bayes_basic}
\end{align}
where $\int p(y^{obs}|\bm{\theta}) \pi(\bm{\theta}) d\bm{\theta}$ is the marginal probability of data, which is often intractable as a (potentially) high-dimensional integral for general models. 

Standard MCMC approaches, e.g. the Metropolis-Hastings (MH) algorithm can address intractable normalizing constants of a posterior density as long as the posterior density of interest is known up to a constant. However, the likelihood itself is only known up to a parameter dependent constant $\psi(\bm{\theta})$, and hence leads to the so-called ``doubly-intractable'' problem, which cannot be dealt with an naive implementation of MH or other conventional MCMC algorithms designed for models with tractable likelihood functions. This gap has led to the development of a body of MCMC approaches that by design generate samples from doubly-intractable posterior densities, most of which rely on augmenting the posterior density so that the augmented posterior probability distribution is easy to sample from. 

The exchange algorithm has evolved as a popular approach for tackling problems with intractable likelihood such as the Ising and Potts models \citep{moller2006efficient, Murray:2006:MDD:3020419.3020463}. The exchange algorithm samples from the augmented distribution, 
\begin{equation}
\label{eq:exchange_algorithm}
    \pi(\bm{\theta}', y', \bm{\theta} | y^{obs}) \propto p(y^{obs}|\bm{\theta}) \pi(\bm{\theta}) q(\bm{\theta}'|\bm{\theta}) p(y'|\bm{\theta}')
\end{equation}
where $p(y^{obs}|\bm{\theta})$ and $p(y'|\bm{\theta}')$ involve the same distributional family but with different parameter values. The distribution $q(\bm{\theta}'|\bm{\theta})$ is any distribution for augmented variable $\bm{\theta}'$ which might depend on $\theta$, for example, a random walk centered at $\bm{\theta}$. Sampling auxiliary variables on an extended state space allows the normalizing constants in likelihood to be cancelled in the Metropolis-Hastings acceptance probability, 
\begin{align}
\label{eq:exchange_algorithm_MH}  
\alpha = & \min \bigg(1, \frac{p(\bm{\theta}') p(y^{obs}|\bm{\theta}') q(\bm{\theta}|\bm{\theta}') p(y'|\bm{\theta})}{p(\bm{\theta}) p(y^{obs}|\bm{\theta}) q(\bm{\theta}'|\bm{\theta}) p(y'|\bm{\theta}')} \bigg) \nonumber \\
= & \min \bigg(1, \frac{p(\bm{\theta}')   q(\bm{\theta}|\bm{\theta}') } {p(\bm{\theta})  q(\bm{\theta}'|\bm{\theta})} \exp \left\{ (\bm{\theta}' - \bm{\theta})^{\intercal} (g(y^{obs}) - g(y')) \right\} \bigg),
\end{align}
which is tractable and therefore the Metropolis-Hastings type algorithm operating on the augmented state space is applicable to general ERGMs by design. However, the exact exchange algorithm requires exact simulation of the auxiliary variable $y'$ from the likelihood, which is typically infeasible for general ERGMs. The \emph{approximate exchange algorithm} (AEA) of \citet{caimo2011bayesian} modifies the original exchange algorithm by substituting MCMC-based approximate samples for exact draws. Specifically, the ``tie-no-tie" (TNT) sampler \citep{morris2008specification} was advocated as a more efficient approach to simulate from ERGM likelihood at each MCMC iteration, according to the implementation in \texttt{bergm} function from the R package \texttt{Bergm} \citep{Bergm}. The default implementation of approximate exchange algorithm in \texttt{Bergm} package uses the idea of adaptive direction sampling (ADS) method \citep{gilks1994adaptive, ter2008differential} from Population Monte Carlo to propose ``parallel ADS" move to improve the mixing, and the default number of chains is set to be twice the number of model parameters.

\section{Approximate Bayesian Computation for ERGMs}
\label{sec:abc}
In this paper, we focus on alternatives to exchange sampling based on kernel methods and approximate Bayesian computation. ABC has emerged as a powerful tool for (approximate) Bayesian analysis of complex models for which the likelihood $p(y|\bm{\theta})$ is unavailable or computationally intractable but where simulation of $Y|\bm{\theta}$ is feasible \citep{pritchard1999population, beaumont2002approximate, marjoram2003markov, sisson2011likelihood, marin2012approximate, sisson2018overview}. 

In approximate Bayesian computation (ABC), inference is concerned with the \emph{partial} posterior distribution $\pi( \bm{\theta} | s^{obs} )$ \citep{doksum1990consistent},
\begin{equation}
\label{eq:partial_posterior}
    \pi(\bm{\theta} | s^{obs} ) = \frac{p(s^{obs})|\bm{\theta} ) \pi(\bm{\theta})}{\int p(s^{obs})|\bm{\theta}) \pi(\bm{\theta}) d\bm{\theta}},
\end{equation}
where $s^{obs} \equiv S(y^{obs})$ represents a vector of $d$-dimensional summary statistics computed from the observed data $y^{obs}$. The classical rejection ABC (R-ABC) (see~Algorithm~\ref{alg:rabc}) approximates the partial posterior distribution $\pi(\bm{\theta} | s^{obs})$ by 

\begin{equation}
\label{eq:ABC_post}
    \pi_{h}^{ABC}(\bm{\theta} | s^{obs}) \propto \int \mathbbm{1}(\norm{s^{*} - s^{obs}} \leqslant  h) p(s^{*}|\bm{\theta} ) \pi(\bm{\theta}) ds^{*}.
\end{equation}

and proceeds by first drawing $N$ values $\bm{\theta}^{(i)}$, $i = 1,\cdots,N$ from the prior distribution $\pi$ and then simulating data from the likelihood $p(y|\bm{\theta}^{(i)})$, retaining those $\bm{\theta}^{(i)}$ with $\norm{s^{(i)} - s^{obs}} \leqslant h$ ($h>0$, usually a sufficiently small number to control the precision of the approximation) under some distance metric $\norm{\cdot}$. The underlying idea (based on the work of \citet{rubin1984bayesianly}) is that $\bm{\theta}^{(i)}$ is unlikely to have generated the observed data, if $\norm{s^{(i)} - s^{obs}}$ is large. Such algorithms converge to the exact posterior when $h\to 0$ and $s$ contains all sufficient statistics, because the posterior $\pi(\bm{\theta} | s^{obs}$) can be regarded as a slice of the joint distribution $\pi(\bm{\theta}, s)$ at $s=s^{obs}$.  A somewhat deeper observation (which we will exploit below) is that R-ABC is a form of \emph{kernel method,} in which a uniform kernel with respect to the metric $\norm{\cdot}$ with bandwidth $h$ used to perform a simulation-based analog of kernel regression (predicting posterior quantities at $s^{obs}$).  This informal intuition \citep[which can be made precise, see e.g.][]{fukumizu2011kernel} suggests a number of potential improvements to the base algorithm, some of which we will leverage here.

 In practice, despite being embarrassingly parallel, a naive implementation of R-ABC can perform poorly given limited computational resources.  In the ERGM context, two immediate problems arise:
\begin{itemize}
    \item Under a weakly informative prior, an extremely large proportion of sampled parameters may generate graphs nowhere close to the observed graph. For example, a prior such as a multivariate Gaussian centered at zero with large standard deviations places most of its mass in unrealistic regions of the parameter space: e.g. positive values of the parameter associated with the \emph{edges} term in ERGMs are rarely seen (when there is no \texttt{edgecov}), as most real-world network data are sparse \citep{kolaczyk2015question}; and large positive values of parameters associated with dependence terms such as \emph{k-stars}, \emph{triangles}, or \emph{shared partners} can lead to \emph{degenerate} probability distribution on graphs that are not useful for network modeling \citep{handcock2003assessing, schweinberger2011instability}.  R-ABC algorithms can be very inefficient under such prior specifications, as the feasible region for realistic real-world networks in the parameter space is often very thin and irregularly-shaped \citep{handcock2003assessing,rinaldo2009geometry}.
    
    \item The distance metric $\norm{\cdot}$ and the rejection threshold $h$ are determined based on the so-called ``reference table'' (simulated parameter-data pairs obtained from a pilot run). The former is usually chosen as a version of weighted Euclidean distance with the weights being selected to normalize the summary statistics so that they vary over roughly the same scale, preventing the distance being dominated by the most variable statistic. The threshold $h$ controls the trade-off between runtime and approximation accuracy, and for R-ABC it is usually selected using the $1 \%$ quantile of the distance computed based on the pilot run.  However, the relatively high cost of ERGM simulation can make such algorithm tuning fairly expensive, especially where sampling must be based on an imprecise prior (which, as described above, will lead in most cases to degenerate or otherwise non-viable graph distributions).
\end{itemize}

\begin{algorithm}
    \caption{Rejection-ABC (R-ABC) algorithm\label{alg:rabc} \citep{pritchard1999population}}
  \begin{algorithmic}[1]
    \REQUIRE Observed summary statistics $s^{obs} = S(y^{obs})$, data generating mechanism $ p(y|\bm{\theta})$, prior $\pi(\bm{\theta})$
    \INPUT Summary statistics $s = S(y)$ \newline
           \indent A desired sample size $N > 0$. \newline
           \indent A distance metric $\norm{\cdot}$ \newline
           \indent A threshold parameter $h>0$. \newline
           \indent Burn-in for MCMC-based simulation for the likelihood (default $B = 2n^2$, where $n$ is the network size) \newline
           \indent Compute observed summary statistics $s^{obs} = S(y^{obs})$
    \WHILE{$i \leqslant N$}
      \STATE $\bm{\theta}' \sim  \pi(\bm{\theta})$
      \STATE $y' \sim p(y|\bm{\theta}')$ (burn-in for the MCMC-based simulation B)
      \STATE $s' = S(y')$
      \IF{$\norm{s' - s^{obs}} \leqslant h$} 
      \STATE Set $\bm{\theta}^{(i)} = \bm{\theta}'$, $i=i+1$ 
      \ENDIF
    \ENDWHILE
     \OUTPUT A set of parameter values $\left\{\bm{\theta}^{(i)}\right\}_{i=1}^{N}$ with equal weights, drawn from  $\pi_{h}^{ABC}(\bm{\theta}| s^{obs})$
  \end{algorithmic}
\end{algorithm}

Given these issues, a number of adaptations are required to make ABC feasible for ERGM inference,  Here, we provide a strategy for approximate Bayesian inference for ERGM parameters under two different scenarios. We first consider the most common scenario, in which we either observe the full network or the set of sufficient statistics is from it, developing a highly parallel algorithm for fast Bayesian inference. In the second scenario, only a subset of the sufficient statistics can be observed (potentially alongside other, proxy statistics), which is typical for sampled, incompletely reported, or obfuscated network data. While conventional estimation schemes are difficult to apply in these cases without extensive re-engineering, we show that our ABC approach easily accommodates them.  In both cases, we propose a version of a kernelized ABC-MCMC algorithm for posterior simulation, though we also discuss KBR-style approaches for efficient posterior moment estimation.

\subsection{Kernel ABC importance sampling algorithm}
\label{subsec:kabc_is}
To improve the sampling efficiency when only weakly informative priors are available, we propose to sample from an importance density rather than the prior.  We also consider alternative kernels to the standard uniform kernel employed in the default R-ABC algorithm.  This leads to a \emph{kernel ABC importance sampling algorithm} (K-ABC-IS), shown here as Algorithm~\ref{alg:kabc_is}. The easy-to-calculate MPLE is a natural choice for constructing an initial proposal distribution when the full network is available. Specifically, we employ a location-scale family centered at the MPLE, with a scale matrix based on the Hessian of the log pseudolikelihood.  Because the curvature of the pseudolikelihood about the MPLE generally underestimates the variability of the parameters, we use an ``inflated'' multivariate Student's t proposal $\mathcal{T}_{\nu}$ with a relatively small degree of freedom parameter (e.g. $\nu=4$) whose scale matrix is the inverse Hessian matrix of the negative log pseudolikelihood at the MPLE, multiplied by a scaling factor $\omega > 1$ to ensure that the sampled parameters are not confined to an overly narrow region near the MPLE. With the parameters sampled from the importance density, we are more likely to generate graphs that are more similar to the observed graph, hence improving sampling efficiency. (Equivalently, we expect that--so long as the data is reasonably informative, and under reasonable choices of priors--substantial posterior mass will be concentrated in the vicinity of the MLE, and hence typically the MPLE.  Use of a heavy-tailed proposal ``hedges'' this expectation against the possibility that the MPLE is a poor initial guess, and ensures adequate weight to majorize the tails of the posterior distribution.)  Algorithm \ref{alg:kabc_is} presents the kernel ABC importance sampling algorithm.  Intuitively, the key idea is to ``doubly re-weight'' the sampled parameters by both their importance ratio versus the prior (i.e., how likely the draw would be to arise under the prior versus the proposal) and their likelihood of generating graphs that are similar to the observed graph in terms of the summary statistics. As a metric on the space of statistics we employ the \emph{Mahalanobis} distance; it serves as a natural choice because it takes both the variability and correlation of the summary statistics into consideration (an important factor, since many typical ERGM statistics are highly correlated).  Although many choices of kernel are possible, we here suggest a Gaussian kernel due to the fact that (1) as a non-compact kernel, it fails gracefully in sparse regions of the simulation space, and (2) it is a fairly efficient estimator for smooth distributions.  The bandwidth, $h$, is here chosen based on a simple heuristic for kernel density estimation \citep{Silverman86} applied to the the computed Mahalanobis distance $d^{(i)}, i=1,\cdots,N$ distribution.  It should be noted that we do not reject samples in this algorithm, instead assigning them different weights according to both importance ratio and kernel weights. We have found that the proposed algorithm \ref{alg:kabc_is} is an improvement over ABC with smooth rejection \citep{beaumont2002approximate} in the ERGM setting.  It is worth noting that more sophisticated approaches for bandwidth selection exist but can increase the computational cost; as discussed in detail in later sections, our experience has suggested that the heuristic bandwith provides comparable performance to more elaborate schemes with much greater computational efficiency.

Note that the sampling step in algorithm \ref{alg:kabc_is} is embarrassingly parallel; since this accounts for the overwhelming majority of the algorithm's computational cost, dramatic performance enhancements are possible on multi-core hardware.  By contrast, the approximate exchange algorithm must be run serially, and cannot take advantage of this level of parallelism.  On the other hand, the exchange algorithm has the advantage of exploring the parameter space in a more controlled manner, guided by the likelihood ratio and prior ratio defined in \eqref{eq:exchange_algorithm_MH} at each iteration, and in our experiments has proved to be slightly more efficient than K-ABC-IS when the latter is run on a single core.  When multiple cores are available, K-ABC-IS can be substantially faster.

\begin{algorithm}
    \caption{K-ABC importance sampling algorithm (K-ABC-IS) \label{alg:kabc_is}}
  \begin{algorithmic}[1]
    \REQUIRE Observed summary statistics $s^{obs} = S(y^{obs})$, data generating mechanism $ p(y|\bm{\theta})$, prior $\pi(\bm{\theta})$
    \INPUT A desired sample size $N > 0$. \newline
           \indent A parametric family for proposal distribution (e.g.multivariate Student's t proposal distribution,  $\mathcal{T}_{\nu}$, degree of freedom $\nu$). \newline
           \indent A scale factor $\omega$ \newline
           \indent Burn-in for MCMC-based simulation for the likelihood (default $B = 2n^2$, where $n$ is the network size) \newline
           \indent A distance metric $\norm{\cdot}$ (e.g. mahalanobis distance) \newline
           \indent Smoothing kernel $K_{h}(\cdot)$ and scale parameter $h>0$. \newline
           \indent (Optional) $\hat{\bm{\theta}}_{MPLE}$, $\bm{I}(\hat{\bm{\theta}}_{MPLE})$
     \STATE \textbf{Initialization}: $\hat{\mu}$, $\hat{\Sigma}$ (default $\hat{\mu} = \hat{\bm{\theta}}_{MPLE}$, $\hat{\Sigma} = \omega \bm{I}^{-1}(\hat{\bm{\theta}}_{MPLE}) )$, \newline
             \indent set $f(\bm{\theta}) \equiv \mathcal{T}_{4}(\hat{\mu},\hat{\Sigma})$
    \FOR{$i = 1,2,\cdots, N$}
      \STATE  $\bm{\theta}^{(i)} \sim  f(\bm{\theta})$, (unnormalized) importance weight $w_{I}^{(i)} = \frac{\pi(\bm{\theta}^{(i)})}{f(\bm{\theta}^{(i)})}$
      \STATE $y^{(i)} \sim p(y|\bm{\theta}^{(i)})$ (burn-in of the MCMC-based simulation $B$)
      \STATE $s^{(i)} = S(y^{(i)})$
    \ENDFOR
    \STATE $W = \frac{1}{N} \sum_{i=1}^{N}(s^{(i)} - \bar{s}) (s^{(i)} - \bar{s})^{\intercal} $, where $\bar{s} = \frac{1}{N} \sum_{i=1}^{N} s^{(i)}$ 
    \STATE $d^{(i)} = (s^{(i)} - s^{obs})^{\intercal} W^{-1} (s^{(i)} - s^{obs})$ for $i=1,2,\cdots,N$
    \STATE Perform univariate kernel density estimate on $d^{(i)}, i=1,2,\cdots,N$ to obtain a (heuristic) bandwidth $h$, and fix $h$ as the scale parameter for the smoothing kernel $K_{h}(\cdot)$, kernel weight $w_{K}^{(i)} \propto K_{h}(d^{(i)})$. 
    \STATE Assign weight to $\bm{\theta}^{(i)}$ as $\tilde{w}^{(i)} \propto w_{I}^{(i)} w_{K}^{(i)}$, and the corresponding normalized $w^{(i)}$
     \OUTPUT A set of parameter values $\left\{\bm{\theta}^{(i)}\right\}_{i=1}^{N}$ with weights $w^{(i)}$, drawn from  $\pi_{h}^{ABC}(\bm{\theta}| s^{obs})$
  \end{algorithmic}
\end{algorithm}

Based on the weighted parameters returned by Algorithm~\ref{alg:kabc_is}, we estimate the partial posterior mean of any scalar function of model parameters, $a(\bm{\theta})$, $\mathbb{E}[ a(\bm{\theta}) | s^{obs}]$ by the kernel estimate
\begin{align}
\label{eq:cond_expectation}
    m^{0,a} = & \sum_{i=1}^{N} w^{(i)} a(\bm{\theta}^{(i)}) \nonumber \\
                                        = &  \frac{\sum_{i=1}^{N} a(\bm{\theta}^{(i)}) w_{I}^{(i)} w_{K}^{(i)}  }{ \sum_{i=1}^{N} w_{I}^{(i)} w_{K}^{(i)} } \nonumber \\                 
                                        = &   \frac{\sum_{i=1}^{N} a(\bm{\theta}^{(i)}) w_{I}^{(i)} K_{h}( d^{(i)} )  }{ \sum_{i=1}^{N} w_{I}^{(i)}  K_{h}( d^{(i)} ) } .
\end{align}
Note that \eqref{eq:cond_expectation} is similar to the Nadaraya-Watson type estimator \citep{nadaraya1964estimating, watson1964smooth}, which can be found by minimizing the weighted sum of squared residuals
\begin{equation}\label{eq:WSSR_ND}
    WSSR_{0} = \sum_{i=1}^{N} \left\{ a(\bm{\theta}^{(i)}) - \alpha \right\}^{2} w^{(i)}
\end{equation}


By letting $a(\bm{\theta}) = \bm{\theta}_{j}$ and $\bm{\theta}_{i} \bm{\theta}_{j}, i,j=1,\cdots,p$, we can obtain the estimate for posterior mean and posterior second moments, hence yielding a natural estimate of posterior variance based on the identity $\Var[ \bm{\theta}_{j} | s^{obs} ] = \mathbb{E}[ \bm{\theta}_{j}^{2} | s^{obs}] - (\mathbb{E}[ \bm{\theta}_{j} | s^{obs}])^{2}$. From a non-parametric regression perspective, the proposed estimator in \eqref{eq:cond_expectation} corresponds to a locally constant estimate, and more intricate estimation of the posterior moments might be achieved by using locally linear or polynomial estimators \citep{blum2010approximate} or any state-of-the-art machine learning techniques as long as the optimization is with respect to the squared error loss (e.g., kernelized WLS). Note that when the sufficient statistics $g(y)$ are a subset of the selected summary statistics, the resulting estimator targets the true posterior mean and standard deviation. 

The construction of posterior intervals is straightforward given weighted samples $\left\{ (\bm{\theta}^{(i)}, w^{(i)} )\right\}_{i=1}^{N} $. For $j=1,\cdots,p$, the general procedure is as follows:
\begin{enumerate}
    \item Find the empirical cumulative distribution function (ECDF) as $\hat{F}_{j}(x) = \frac{1}{N} \sum_{i=1}^{N} \mathbbm{1}( \bm{\theta}_{j}^{(i)} \leqslant x)$.
    \item Obtain a smooth approximation $\tilde{F}_{j}(x)$ of the ECDF $\hat{F}_{j}(x)$ using monotonic spline (e.g. \texttt{splinefun} function in R, with method set as ``monoH.FC").
    \item Find the $q$-th quantile of $\tilde{F}_{j}(x)$ by minimizing the squared error.
\end{enumerate}
To obtain independent samples from the joint posterior with equal weights, it is possible to use sampling-importance resampling (SIR) techniques \citep{rubin1987comment, rubin1988using}. The resampling step can be conducted with or without replacement, but the latter should be favored when only a few large weights and many small weights are present \citep{gelman:etal:1995}. Improved SIR with faster convergence rates and bias-reduced SIR were proposed and studied by \citet{skare2003improved}. The SIR-based techniques have proved to be useful for ABC algorithms producing weighted samples \citep[see, e.g.][]{mengersen2013bayesian, zhu2016bootstrap}. 

The theoretical validity of the proposed ABC algorithm can be justified from an approximate likelihood perspective \citep{karabatsos2018approximate}, as it implicitly works with a kernel density estimate of the likelihood, i.e.
\begin{equation}
\label{eq:ABC_post}
    \pi_{h}^{ABC}(\bm{\theta} | s^{obs}) \propto \int K_{h}(\norm{ s^{*} - s^{obs} }) p(s^{*}|\bm{\theta}) \pi(\bm{\theta}) ds^{*}.
\end{equation}
In particular, in the typical ERGM case $s$ corresponds to the sufficient statistics of the proposed model, and hence $\pi_{h}^{ABC}$ can closely approximate the true posterior for appropriate choice of $K_h$.  As we show below, a Gaussian kernel appears to work well for the cases studied here.

\subsection{Kernel ABC adaptive importance sampling algorithm}
\label{subsec:kabc_ais}
We consider the idea of adaptive importance sampling (AIS) \citep[see,e.g.][]{ortiz2000adaptive,Liu:2008:MCS:1571802, pennanen2006adaptive, Rubinstein:2004:CEM:1014902} where the initial proposal distribution is not good enough. This idea can be particularly useful when MPLE is suspected to be heavily biased (e.g. network size is small and with strong dyadic dependence) or even not available (e.g. fitting egocentrically sampled data with terms involving counts of triangles or higher-order cycles). Algorithm~\ref{alg:kabc_ais} describes the K-ABC adaptive importance sampling (K-ABC-AIS) algorithm, in which both the proposal distribution and distance function are updated iteratively based on the points sampled in most recent step. \citet{prangle2017adapting} gave some theoretical support for similar algorithms with compact smoothing kernels. 

\begin{algorithm}
    \caption{K-ABC adaptive importance sampling algorithm (K-ABC-AIS) \label{alg:kabc_ais}}
  \begin{algorithmic}[1]
    \REQUIRE Observed summary statistics $s^{obs} = S(y^{obs})$, data generating mechanism $ p(y|\bm{\theta})$, prior $\pi(\bm{\theta})$
    \INPUT  $K_{h}(\cdot)$ ($h>0$), B, distance metric $\norm{\cdot}$ (e.g. mahalanobis distance), $\mathcal{T}_{\nu}$  \newline
           \indent Number of rounds of importance sampling $T$. \newline
           \indent Desired sample sizes $N_{t} > 0, t=1,\cdots,T$. \newline
           \indent Scale factors $\omega_{t} > 0, t=1,\cdots,T$ \newline
            \indent (Optional) $\hat{\bm{\theta}}_{MPLE}$, $\bm{I}(\hat{\bm{\theta}}_{MPLE})$
    \STATE \textbf{Initialization}: Let $t=1$ and find $\hat{\mu}_{1}$, $\hat{\Sigma}_{1}$ (default $\hat{\mu}_{1} = \hat{\bm{\theta}}_{MPLE}$, $\hat{\Sigma}_{1} = \omega_{1} \bm{I}^{-1}(\hat{\bm{\theta}}_{MPLE}) )$, \newline
             \indent set $f_{1}(\bm{\theta}) \equiv \mathcal{T}_{\nu}(\hat{\mu}_{1},\hat{\Sigma}_{1})$
    \FOR{$t=1,\cdots,T$}
     \FOR{$i=1,\cdots,N$}
      \STATE $\bm{\theta}_{t}^{(i)} \sim  f_{t}(\bm{\theta})$, (unnormalized) importance weight $w_{I,t}^{(i)} = \frac{\pi(\bm{\theta}_{t}^{(i)})}{f(\bm{\theta}_{t}^{(i)})}$
      \STATE $y_{t}^{(i)} \sim p(y|\bm{\theta}_{t}^{(i)})$ (burn-in of the MCMC-based simulation $B$)
      \STATE $s_{t}^{(i)} = S(y_{t}^{(i)})$
     \ENDFOR
    \STATE $W_{t} = \frac{1}{N} \sum_{i=1}^{N}(s_{t}^{(i)} - \bar{s}) (s_{t}^{(i)} - \bar{s})^{\intercal} $, where $\bar{s}_{t} = \frac{1}{N} \sum_{i=1}^{N} s_{t}^{(i)}$ 
    \STATE $d_{t}^{(i)} = (s_{t}^{(i)} - s^{obs})^{\intercal} W_{t}^{-1} (s_{t}^{(i)} - s^{obs})$ for $i=1,\cdots,N$
    \STATE Perform univariate kernel density estimate on $d_{t}^{(i)}, i=1,2,\cdots,N$ to obtain a (heuristic) bandwidth $h_{t}$, and fix $h_{t}$ as the scale parameter for the smoothing kernel $K_{h_{t}}(\cdot)$ 
    \STATE Assign weights to $\bm{\theta}_{t}^{(i)}$ as $\tilde{w}_{t}^{(i)} \propto w_{I,t}^{(i)} w_{K,t}^{(i)}$, and the corresponding normalized $w_{t}^{(i)}$
    \STATE $\hat{\mu}_{t} = \sum_{i=1}^{N_{t}} w_{t}^{(i)}\bm{\theta}_{t}^{(i)}$, $\hat{\Sigma}_{t} = \omega_{t} \sum_{i=1}^{N_{t}} w_{t}^{(i)} (\bm{\theta}_{t}^{(i)} - \hat{\mu}_{t}) (\bm{\theta}_{t}^{(i)} - \hat{\mu}_{t})^{\intercal}$.
    \IF{$t \textless T$}
    \STATE $f_{t+1}(\bm{\theta}) \equiv \mathcal{T}_{4}(\hat{\mu}_{t},\hat{\Sigma}_{t})$
    \ENDIF
     \ENDFOR
     \OUTPUT A set of $i=1,\cdots,N_{T}$ samples $\left\{\bm{\theta}^{(i)}\right\}_{i=1}^{N_{T}}$ with weights $w_{T}^{(i)}$, drawn from $\pi_{h}^{ABC}(\bm{\theta}| s^{obs})$
  \end{algorithmic}
\end{algorithm}


\subsection{Proposal Distributions for Importance Sampling}
\label{subsec:prop_dist}
Similar to the importance of proposal distributions in MCMC \citep{roberts1997weak,rosenthal2011optimal}, our proposed algorithms can greatly benefit from a well-chosen proposal distribution. Our focus here is on probability densities constructed from a common yet flexible distributional family, the multivariate Student's t distribution, $\mathcal{T}_{\nu}(\mu, \Sigma)$. The easy-to-calculate MPLE is a natural choice for $\mu$, as it is typically not very far from the high density region (or, at minimum, likely to be closer than the prior mean). To mitigate the potential issue caused by an overly confident estimate of uncertainty, we consider a relatively small degree of freedom $\nu = 4$ and use a scale factor $\omega=4$ to inflate the nominal covariance matrix given by MPLE. When using adaptive importance sampling, we advocate the use of a sequence of gradually decreasing scaling factors $\omega_{1}, \cdots, \omega_{T}$ so as to avoid wasting too many draws in low-density regions. Similarly, increasing the degree of freedom in later rounds of importance sampling is also an option.    
One potential pitfall for K-ABC-AIS is to split the fixed computational budget into multiple, very thin portions, which can in turn lead to an even worse proposal distribution than the one initially suggested (e.g., if the first round yields an estimate that is worse than the MPLE itself due to insufficient sampling).  We note that samples from previous rounds can be retained in subsequent calculations, provided that their importance weights are handled appropriately; otherwise, however, our observation has been that using more than two to three rounds of refinement yields little benefit, and hence it is more efficient to split a fixed sampling budget into two (or at most three) waves of sampling than in a larger number.  We describe the results of a simulation experiment investigating the impact of sample size below.


\subsection{Bandwidth Selection}
\label{subsec:bw_selection}
A key parameter determining the accuracy of inference via ABC is the bandwidth, $h$. If $s$ is sufficient for $\bm{\theta}$, and $h \rightarrow 0$ we can obtain an arbitrarily good approximation to the true posterior; however, this insight is of relatively little practical use, since exact matching of simulated to observed statistics is an event of vanishingly small probability. A good working bandwidth thus strikes a balance between accuracy in approximating the target distribution (or at least its first several moments) and computational efficiency. As introduced in \ref{subsec:kabc_is}, we find that a simple bandwidth heuristic calculated on the distribution of the simulated Mahalanobis distances can yield satisfactory and stable performance at very low cost. To achieve a higher accuracy on the estimation of posterior moments, alternative approaches that might better serve the purpose in principle include \emph{cross validation} (CV), or k-nearest neighbor CV (kNN CV), given that the goal is to estimate  $\mathbb{E}[ a(\bm{\theta}_{j}) | s^{obs}], j=1,\cdots,p$ at the observed summary statistics $s^{obs}$.  Bandwidths can also be chosen for each dimension, albeit with modification of the kernel and distance metric.  However, preliminary experiments using these methods suggested that bandwidth selection using these approaches was often unstable, and did not yield systematic improvement on the heuristic option.  At the same time, these methods were substantially more computationally expensive than the heuristic, increasing estimation time.  Per-statistic bandwidth selection methods also create challenges for tasks requiring the same weight to be applied to all elements of a draw (e.g., posterior sampling, as opposed e.g. to estimation of marginal posterior moments), as orthogonal kernels lose the advantage of the Mahalanobis distance in accounting for correlations among statistics (and thereby efficiency) and correlated kernels are difficult to calibrate.  Because we found more sophisticated bandwidth selection schemes to add cost without improving performance, we do not pursue them further here.  However, it is plausible that better procedures are possible, and we regard this as an open problem.


\section{Applications}
\label{sec:applications}
We apply our approach to two benchmark social network datasets of varying sizes. Approximate exchange algorithm (AEA) was considered to be the current "gold-standard" for Bayesian inference of ERGMs. Hence, to illustrate the accuracy and computational efficiency of our approach, we compare the proposed algorithm with AEA. All computations in this paper are implemented in \textbf{R} \citep{R-Team} on a computing server (96GB RAM, with 4 Intel Xeon E5-2690v2 processors, operating at 3.00GHz, with 10 processing cores in each) -- we use software suite \texttt{statnet} \citep{handcock2008statnet} to simulate networks from ERGMs, and we implement AEA using the R package \texttt{Bergm} \citep{Bergm}. (Note that \texttt{Bergm} also uses \texttt{statnet} for MCMC simulation, and hence its implementation and those of K-ABC methods employed in this paper are directly comparable.)  The \textbf{R} code to implement the algorithm and the data are available from \url{https://github.com/fyin-stats/K_ABC_ERGMs}.


\subsection{Karate Club network}
\label{subsec:karate_club}
\begin{figure}
\centering
\includegraphics{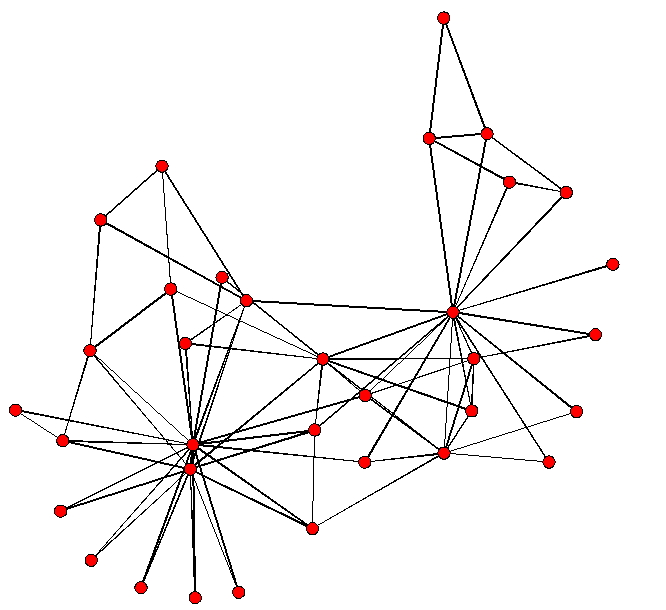}
\caption{Karate club friendship network \label{pic:karate}}
\end{figure}

The Karate club data \citep{zachary1977information} represents a friendship network between 34 members in a US university karate club in the 1970's. This network consists of 78 undirected edges as presented in Figure \ref{pic:karate}, and the interest lies on the effect of triad closure. We consider the optimal model specification identified in \citet{bouranis2018bayesian}, which is $g(y) = (g_{1}(y), v(y,\phi) )$. Specifically, $g_{1}(y) = \sum_{i<j} y_{ij}$ is the total number of edges in the network and $v(y,\phi)$ is the \emph{geometrically weighted edgewise shared partner}(GWESP) statistic \citep{hunter2006inference} defined as

$$ v(y,\phi) = e^{\phi} \sum_{k=1}^{n-2} \left\{ 1 - (1-e^{-\phi})^{k}   \right\} EP_{k}(y)  $$

where $EP_{k}(y)$ is the number of connected pairs that have exactly $k$ common neighbors and parameter $\phi$ controls the decreasing rates of weights placed on higher order terms. The GWESP statistic is a common choice for modeling the tendency of forming local clusters in a network, and it has intuitive interpretation as there is diminishing positive return on the odds of an edge for each additional shared partner (e.g. one more common friend in the context of friendship network). 

We obtain the ``ground truth'' based on a long AEA run consisting of $4$ population chains (burn-in period $2500$, $12500$ main iterations for each chain) with a ``conservative" burn-in for MCMC-based simulation equal to $100000$, which takes 4373.7 seconds (1.215 hours) to fit. While acknowledging that a holistic comparison between K-ABC and AEA sampler cannot be easily conducted due to algorithmic differences, we here provide a more limited comparison of AEA versus the K-ABC approach for typcal desiderata within a basic test case. When the total size of proposed samples is fixed, prior theory leads us to expect the AEA to outperform K-ABC. However, K-ABC is embarrassingly parallel, and therefore can base inference on sample sizes that scale with the number of available cores, given fixed computational time. (Equivalently, given a fixed number of simulated graphs, wallclock time can be reduced under K-ABC by employing a larger number of cores.)  Taking these facts into consideration, we consider the following settings,

\begin{itemize}
    \item K-ABC-IS : One round of importance sampling, sample size: 32000, degree of freedom $\nu = 4$, scale factor $\omega = 4$, burn-in for MCMC-based simulation $B = 10^4$.
    \item K-ABC-AIS : Two rounds of importance sampling, sample size: (8000,24000), degree of freedom $\nu_{1} = 4, \nu_{2} = 4$, scale factor $\omega_{1} = 4, \omega_{2} = 2 $, burn-in for MCMC-based simulation $B = 10^4$.
    \item AEA : $4$ population chains, each chain with burn in = 500, main iters = 1500, auxiliary burn-in (i.e. burn-in for MCMC-based simulation) = $10^4$ 
\end{itemize}

Note that we allow the K-ABC-IS and K-ABC-AIS to draw a total of $32000$ samples, which is $4$ times the total sample size in AEA. To ensure the simulated networks used in these three algorithms are of the same quality, we fix the burn-in period of MCMC-based simulation to be $10000$. Under the above settings, all the algorithms are run $20$ times, and their results are summarized in Table \ref{tb:karate_kbergm_bergm}. Given the stochastic nature of these algorithms, the results differ from run to run, but overall they are very close to the ground truth. K-ABC-AIS and AEA yield essentially identical performance with respect to the estimation of posterior means, but the runtime of K-ABC-AIS is almost one seventh of that for AEA. We also note that the K-ABC-AIS performs better than K-ABC-IS, as the former gives more accurate posterior mean estimates, indicating that the adaptive scheme is indeed helpful for producing a better proposal distribution.

\begin{figure*}
\centering
\includegraphics[width=\textwidth]{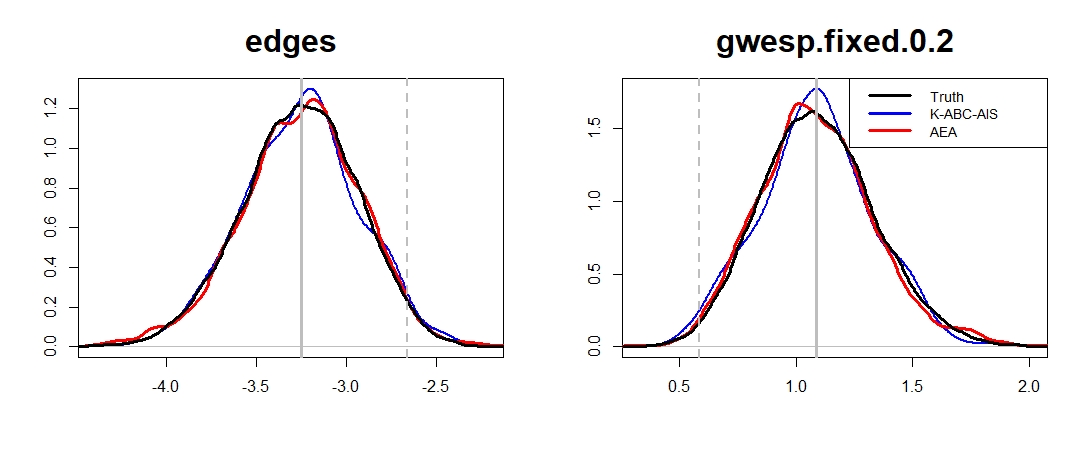}
\caption{Estimated marginal posterior distribution of $\theta$. The grey line and grey dotted line represent the MLE and MPLE, respectively. \label{pic:karate_kbergm_bergm_distn}}
\end{figure*}

Figure \ref{pic:karate_kbergm_bergm_distn} shows that K-ABC matches closely to each marginal posterior distribution from the ground truth. It is worth mentioning that the posterior marginal density for K-ABC is constructed based on unweighted samples obtained using sampling-importance resampling (SIR) with replacement. 

\begin{table}
\centering
\caption{Comparison between K-ABC and AEA. K-ABC algorithms (K-ABC-IS, K-ABC-AIS) are run on 30 cores. Wall-clock runtime reported is the average across 20 runs. \label{tb:karate_kbergm_bergm}}
\resizebox{0.5\textwidth}{!}{\begin{tabular}{lllll}
  \hline

 & Ground truth & MAE & RMSE & Runtime (secs)  \\
  \hline
   K-ABC-IS ($\theta_{1}$) & -3.25 & 0.09 & 0.11 & 14.2   \\
   K-ABC-IS ($\theta_{2}$) & 1.10 & 0.07 & 0.08 &  14.2  \\
  \hline
  K-ABC-AIS ($\theta_{1}$) & -3.25 & 0.03 & 0.03 & 14.8  \\
  K-ABC-AIS ($\theta_{2}$) & 1.10 & 0.02 & 0.03 & 14.8  \\
  \hline
  AEA ($\theta_{1}$) & -3.25 & 0.02 & 0.03 & 94.4  \\
  AEA ($\theta_{2}$) & 1.10 & 0.02 & 0.02 & 94.4 \\
   \hline
\end{tabular}}
\end{table}

\subsection{Faux Mesa High School network}
\label{subsuc:faux_mesa}
\begin{figure}
\centering
\includegraphics[width=0.5\textwidth]{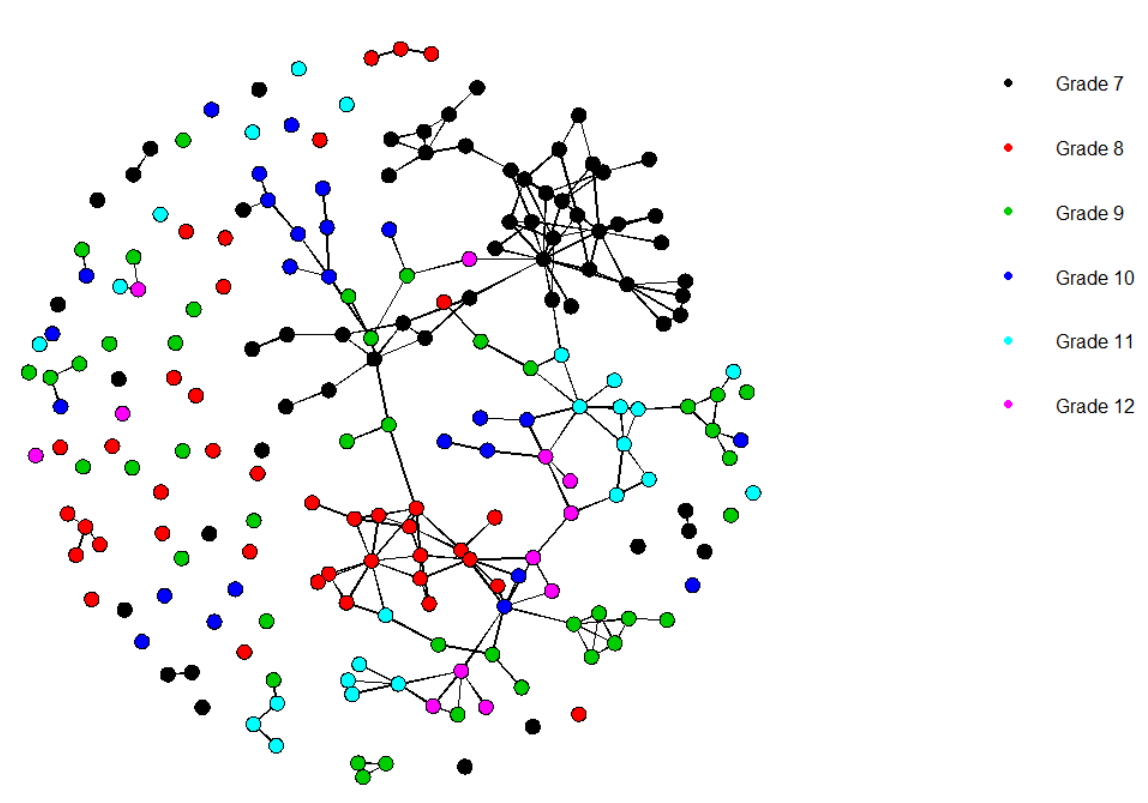}
\caption{Faux Mesa High School friendship network. Colours indicate the grade. \label{pic:faux}}
\end{figure}

The Faux Mesa High school network represents a total of 203 undirected friendship relations in a synthetic high school of 205 students based on an observed high school in the western U.S. \citep{handcock2008statnet}, and it is widely used as a realistic test network for statistical procedures. Figure \ref{pic:faux} shows that the network is sparse and a large proportion of edges are formed between students in the same grade, suggesting a strong \emph{homophily} effect on grade. The presence of some local clusters is indicative of the bias towards the formation of triangles (i.e. transitivity effect). Bearing the observed facts in mind, we consider a model with the following 3 statistics: 

$$ g_{1}(y) = \sum_{i<j} y_{ij} \ \ \ \ \ \ g_{2}(y) = \sum_{i < j} y_{ij} \mathbbm{1}( x_{i} = x_{j} )  $$

$$ g_{3}(y) = v(y, \phi)$$

where $x_{i}$ represents the grade of $i$-th individual and $\mathbbm{1}(\cdot)$ is the indicator function, hence $g_{2}(y)$ counts the total number of edges connecting individuals from the same grade. $v(y, \phi)$ is the \emph{geometrically weighted edgewise shared partner} (GWESP) statistic

$$ w(y, \phi) = e^{\phi} \sum_{k=1}^{n-2} \left\{ 1 - (1-e^{-\phi})^{k}   \right\} EP_{k}(y)$$

where the decay parameter is fixed at $0.5$ here, as suggested in the model proposed in \citet{hunter2008ergm}.

As large friendship networks are usually sparse with a high degree of homophily and transitivity \citep{goodreau2007advances}, we consider a multivariate Gaussian prior centered at $\mu_{0}=(-2,0.5,0.5)$ and covariance matrix $\Sigma_{0} = 5 \bm{I}_{3}$; this corresponds to an \emph{a priori} belief that the coefficient associated with the edges term $g_{1}(y)$ is likely to be negative and those associated with the effect of grade homophily and transitivity are likely to be positive. The relatively large standard deviations ensure the statistical inference cannot be dominated by the prior belief. Also, given the sparsity of the observed data (network density $\approx$ 0.01), the observed edge-count based sufficient statistics might be a lot closer to their lower bound (0) than to their upper bound (total number of edges, $\frac{n(n-1)}{2}$) and their distribution might be right-skewed, hence the proposed weighting scheme will unfairly favor the sparse graph. As a remedy, we consider a monotonic power-law transformation, $T(u) = \sqrt{u+1}$ on the sufficient statistics when implementing K-ABC procedure. 

The ``ground truth'' is again obtained based on a ``conservative'' AEA run, where we choose a large number of auxiliary iterations, e.g. $5 \times 10^5$, and run AEA for sufficiently long -- 6 population chains, each with burn-in = $4000$, main-iters = $16000$ to ensure that the resulting samples can provide an adequate approximation to the ``true'' target ($17.7$ hours, 63669.7 seconds). We compare the K-ABC-AIS algorithm with the AEA under the following settings --

\begin{itemize}
    \item K-ABC-AIS : Two rounds of importance sampling, sample size: (24000,96000); degree of freedom $\nu_{1} = 4, \nu_{2} = 4$; scale factor $\omega_{1} = 4, \omega_{2} = 2 $, burn-in for MCMC-based simulation $B = 5 \times 10^4$.
    \item AEA : $6$ population chains, each chain with burn-in = 1000, main iters = 4000, auxiliary burn-in (i.e. burn-in for MCMC-based simulation) = $5 \times 10^4$ 
\end{itemize}

\begin{figure*}
\centering
\includegraphics[width=\textwidth]{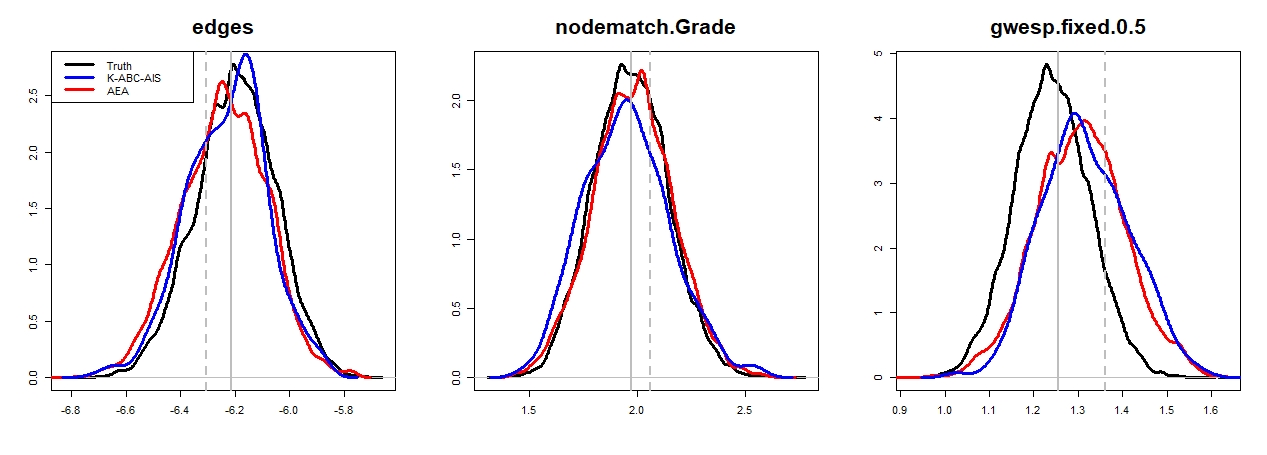}
\caption{Estimated marginal posterior distribution of $\theta$. The grey line and grey dotted line represent the MLE and MPLE, respectively. \label{pic:faux_kbergm_bergm_distn}}
\end{figure*}

Note that in this case we also allow the K-ABC-AIS to draw a total of $120000$ samples, which is $4$ times the total sample size in AEA. Table \ref{tb:faux_kbergm_bergm} shows that the point estimates given by K-ABC-AIS and AEA show virtually identical performance. Figure \ref{pic:faux_kbergm_bergm_distn} shows that the estimated marginal distributions are similar, but we also notice that there is discrepancy between the marginal posterior distribution of the GWESP parameter estimated based on the K-ABC-AIS, AEA and the ``ground truth.'' Such behavior suggests that a sufficiently long burn-in period for simulating from ERGMs can play a crucial role in both AEA and K-ABC type algorithms. 

\begin{table}
\centering
\caption{Comparison between K-ABC and AEA. K-ABC algorithm (K-ABC-AIS) is run on 30 cores. Wall-clock runtime reported is the average across 20 runs. \label{tb:faux_kbergm_bergm}}
\resizebox{0.5\textwidth}{!}{\begin{tabular}{lllll}
  \hline
  & Ground truth & MAE & RMSE & Runtime (secs) \\
  \hline
  K-ABC-AIS ($\theta_{1}$) & -6.20 & 0.04 & 0.05 & 283.6 \\
  K-ABC-AIS ($\theta_{2}$) & 1.97 & 0.01 & 0.01 & 283.6 \\
  K-ABC-AIS ($\theta_{3}$) & 1.24 & 0.07 & 0.07 & 283.6 \\
  \hline
  AEA ($\theta_{1}$) & -6.20 & 0.03 & 0.03 & 1078.3 \\
  AEA ($\theta_{2}$) & 1.97 & 0.01 & 0.01 & 1078.3 \\
   AEA ($\theta_{3}$) & 1.24 & 0.06 & 0.06 & 1078.3 \\
   \hline
\end{tabular}}
\end{table}

\subsection{Computational efficiency of K-ABC with parallel computing}
\begin{figure*}
\centering
\includegraphics[width=\textwidth]{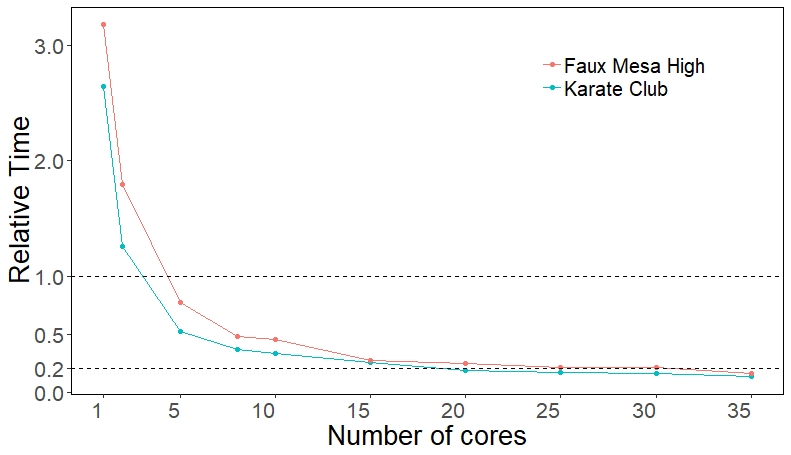}
\caption{The y-axis gives the ratio of the K-ABC-AIS time to that of the AEA time. Values below 1 indicate that the K-ABC-AIS requires a shorter computing time.  \label{pic:relative_time}}
\end{figure*}

We further investigate the computational efficiency of the proposed K-ABC approach. The presented results suggest that - (1) K-ABC seems to be able to produce comparable results to AEA when the total sample size is $4$ times that of AEA; (2) K-ABC-AIS can produce more accurate estimations than K-ABC-IS, and it is advisable to allocate one fifth of the total sample size to the first round of importance sampling step. Therefore, we compare the computational efficiency between K-ABC-AIS and AEA under the settings which give similar level of estimation accuracy. Figure \ref{pic:relative_time} illustrates the relative computing time of the K-ABC-AIS algorithm and AEA for the two networks Karate club (34 nodes), Faux Mesa High (205 nodes) for an increasing number of computing cores. 


The relative computing time is defined as the ratio of K-ABC-AIS time to AEA time, and thus a relative computing time greater than 1 indicates that the AEA computing time is shorter, while a relative computing time smaller than 1 indicates that the K-ABC-AIS provides faster results. 

Figure \ref{pic:relative_time} demonstrates that both networks only require five cores for the K-ABC-AIS to outperform the computing time of the AEA and that the computing time can be further reduced if more computing cores are available as we can get five-fold reduction on the computing time when 30 computing cores are used. We expect further reduction on the computing time as the serial part of the K-ABC-AIS algorithm only takes a small portion of the total runtime.

\section{Further Extensions}
\label{sec:externsions}
The proposed K-ABC approach has a wide range of connections to existing Bayesian computation techniques, including regression-adjustment ABC, Bayes Linear Analysis, and Kernel Bayes' rule.  Techniques and extensions developed in these literatures could naturally be applied our case, without requiring extensive modification of our approach.

It is particularly worth noting the connection between K-ABC and Kernel Bayes' rule (KBR) \citep{fukumizu2011kernel}. Both of them provide posterior estimates in the form of a kernel mean, but the fundamental goal of K-ABC is obtaining samples from an approximation to the posterior distribution, while KBR can generate empirical estimates via the kernel approaches that converge to the true posterior mean embedding in the limit of infinite sample size \citep{fukumizu2013kernel}.  The connection with KBR also makes plain the extent to which estimation of posterior moments (and hence quantiles) is fundamentally a nonparametric regression problem, where we seek to estimate $\mathbb{E}[a(\bm{\theta}) | y^{obs}]$ (for some function $a(\cdot)$) from a superpopulation defined by the joint distribution of $y$ and $\bm{\theta}$.  Because this regression is for us a purely computational device, any scheme that performs well and is computationally efficient is potentially useful.  While we here use an approach that is equivalent to classical kernel regression, kernelized weighted least squares would be another natural choice, as might more exotic alternatives such as random forests or neural networks.  The primary advantage of such methods is their flexibility in fitting complex functions with minimal user input, an asset that is of obvious relevance in this application.  On the other hand, methods that require expensive training procedures to calibrate nuisance parameters may not improve performance sufficiently to justify the increased cost.  Further work will need to be done to determine which techniques, including linear adjustments \citep{beaumont2002approximate, blum2013comparative}, non-linear adjustments \citep{blum2010non}, yield net performance gains.

Another possible direction might be approximating the posterior by a multivariate normal distribution, based on the classic Bernstein-von Mises theorem \citep{van2000asymptotic}. There is recent work on variational Bayesian inference for ERGMs \citep{tan2020bayesian} based on the adjusted pseudo-likelihood \citep{bouranis2018bayesian}, in which the posterior density $\pi(\bm{\theta} | y^{obs})$ is approximated by a Gaussian distribution $q_{\lambda}(\bm{\theta})$, and the parameters $\lambda = \left\{ \mu, \Sigma\right\}$ are found by minimizing the Kullback-Leibler divergence (or equivalently maximizing the \emph{evidence lower bound}). Provided the Gaussian approximation is valid, the proposed approach enables the construction of Gaussian distribution based on estimated posterior mean and posterior second moments. In case the resulting covariance matrix is not positive definite, post correction methods can be adopted \citep[e.g.,][]{loland2013statistical}. To date, general variational inference for ERGMs without the knowledge of MLE has not proven successful outside of demonstration models, and it is unclear whether its limitations can be overcome. However, variational approximations may be useful as additional tools for seeding ABC proposal distributions, especially in the dense graph regime where typical MCMC algorithms are often slow.

Finally, we note that extensions of the methods considered here to temporal ERGMs (TERGMs), ERGMs with latent variables, or other more complex cases are fairly straightforward given the ability to simulate from the data generating process.  In particular, the modular structure of the K-ABC algorithm makes it relatively easy to accommodate such extensions within a single computational framework, so long as a simulation algorithm for the extended model is available.  This is in contrast with existing strategies for ERGM inference, where are generally specialized for fairly narrow classes of models.  This feature makes K-ABC a promising foundation for building ERGM-based modeling tools that are substantially more flexible than those currently in use.

\section{Conclusion}
\label{sec:conc}
In this paper, we introduced a kernelized approximate Bayesian computation (K-ABC) procedure for ERGMs, exploiting the algorithm's parallelizability to show substantial performance gains versus standard methods when multiple cores are available.  In typical cases, the availability of sufficient statistics facilitates inference using this approach, as does the availability of relatively inexpensive crude initial estimates that can be used to construct effective proposal distributions for importance sampling.  Further enhancements in performance can be obtained by iterative refinement of initial estimates, though our simulation studies suggest that (given a fixed budget) a small number of larger samples is usually preferred to many waves of small samples.  Comparing our approach with the current state-of-the-art (the approximate exchange algorithm), we find that K-ABC adaptive importance sampling algorithm (K-ABC-AIS) is able to produce estimates of comparable quality at greatly reduced wall-clock time so long as multiple cores are available.  In a serial setting (i.e., when only one core can be used), the more refined sampling scheme of AEA is more efficient than the ABC techniques explored here, and we would recommend it as the preferred approach in this case.  AEA also has the advantage of providing very high-quality posterior approximations when run with sufficiently rigorous settings (albeit at very high cost).  The two approaches thus have distinct advantages and disadvantages.  One potentially useful asset of the proposed K-ABC algorithm is that it is immediately extensible to non-standard cases (such as inference from proxy statistics) that are difficult to handle using other techniques.  It is also far easier to implement than AEA.  This makes K-ABC a natural choice when flexibility or ease of implementation are considerations, especially if speed is of the essence. 

Though parsimoniously modeling dependencies of scientific interest in networks is the primary objective for ERGMs, the development of efficient Bayesian inference on higher-dimensional ERGMs is favorable. With the recent development on high dimensional ABC algorithms (\citep{nott2014approximate, li2017extending}), we envision ABC as a promising framework.

Finally, we note that there are many variations on the specific implementation decisions pursued here; though we investigated the consequences of several such decisions via two case studies, there are far more possibilities for expansion and modification of the base algorithm than can be considered in any one study.  We are thus optimistic for the potential for further enhancement of this very promising approach to ERGM inference.

\bibliography{mybibfile}


\end{document}